\documentclass[%
 preprint,
superscriptaddress,
 amsmath,amssymb,
 aps,
]{revtex4-1}

\usepackage{graphicx}% Include figure files
\usepackage{dcolumn}% Align table columns on decimal point
\usepackage{bm}% bold math
\usepackage{amsfonts}
\usepackage{hyperref}% add hypertext capabilities
\usepackage[mathlines]{lineno}% Enable numbering of text and display math
\newcommand{\snn}{\sqrt{s_\text{NN}}}

\begin{document}

%\preprint{}

\title{Study of QCD critical point with three-nucleon correlations in light nuclei yields ratios using PYTHIA8/Angantyr}

\author{Zuman Zhang}\thanks{Email: zuman.zhang@hue.edu.cn}
\affiliation{School of Physics and Mechanical Electrical \& Engineering, Hubei University of Education, Wuhan 430205, China}
\affiliation{Institute of Theoretical Physics, Hubei University of Education, Wuhan 430205, China}
\affiliation{Key Laboratory of Quark and Lepton Physics (MOE), Central China Normal University, Wuhan 430079, China}
\author{Ning Yu}\thanks{Email: ning.yuchina@gmail.com(corresponding author)}
\affiliation{School of Physics and Mechanical Electrical \& Engineering, Hubei University of Education, Wuhan 430205, China}
\affiliation{Institute of Theoretical Physics, Hubei University of Education, Wuhan 430205, China}
\affiliation{Key Laboratory of Quark and Lepton Physics (MOE), Central China Normal University, Wuhan 430079, China}
\author{Sha Li}\thanks{Email: lisha@hue.edu.cn}
\affiliation{School of Physics and Mechanical Electrical \& Engineering, Hubei University of Education, Wuhan 430205, China}
\affiliation{Institute of Theoretical Physics, Hubei University of Education, Wuhan 430205, China}
\author{Shuang Li}\thanks{Email: lish@ctgu.edu.cn}
\affiliation{College of Science, China Three Gorges University, Yichang 443002, China}
\affiliation{Center for Astronomy and Space Sciences, China Three Gorges University, Yichang 443002, China}
\author{Siyu Tang}\thanks{Email: tsy@wtu.edu.cn}
\affiliation{School of Mathematical \& Physical Sciences, Wuhan Textile University, Wuhan 430200, China}
\author{Meimei Zhang}\thanks{Email: 18271979598@163.com}
\affiliation{School of Physics and Mechanical Electrical \& Engineering, Hubei University of Education, Wuhan 430205, China}
 \affiliation{College of Science, China Three Gorges University, Yichang 443002, China}

\date{\today}% It is always \today, today,

\begin{abstract}
This study utilizes the PYTHIA8 Angantyr model to systematically investigate the effects of three nucleons correlation $C_{n^2p}$ on the light nuclei yield ratio $N_tN_p/N_d^2$ in Au+Au collisions at $\snn$ = 7.7, 11.5, 14.5, 19.6, 27, 39, 62.4, and 200 GeV. The analysis explores this property across different rapidity ranges, collision centralities, and collision energies, while also examining the roles of multi-parton interactions (MPI) and color reconnection (CR) mechanisms. The results show that the light nuclei yield ratio remains stable with changes in rapidity coverage and collision centrality but slightly increases with rising collision energy. The impact of CR on the light nuclei yield ratio depends entirely on the presence of MPI; when MPI is turned off, CR has no effect. Additionally, the three-nucleon correlation, enhances the light nuclei yield ratio in both central and peripheral collisions. However, the non-monotonic energy dependence observed in experiments, the peak at $\snn=20\sim30$ GeV reported by the STAR experiment, cannot be explained by the Angantyr model due to its lack of key mechanisms related to the quark-gluon plasma (QGP). Nevertheless, the Angantyr model serves as an important baseline for studying collision behaviors in the absence of QGP effects.

\begin{description}\item[PACS numbers]
\verb+25.75.-q, 25.75.Nq+
\end{description}
\end{abstract}

\maketitle

\section{Introduction}
\label{sec:intro}

  The creation of a state of matter known as the quark-gluon plasma (QGP), characterized by deconfined quarks and gluons, is believed to occur under extreme conditions of temperature and/or density during heavy-ion collisions at ultra-relativistic energies. Understanding the phase structure of strongly interacting matter described by Quantum Chromodynamics (QCD) is a key objective in nuclear physics. The QCD phase diagram, typically represented in a two-dimensional graph of temperature ($T$) versus baryon chemical potential ($\mu_{\mathrm{B}}$), offers critical insights into the behavior of strongly interacting matter.

  Lattice QCD calculations have shown that the transition from the hadronic phase to the QGP at low values of $\mu_{\mathrm{B}}$ occurs as a smooth crossover~\cite{ref:QCDphaseplot,Aoki:2006we}. However, at higher values of $\mu_{\mathrm{B}}$, QCD-based model calculations predict a first-order phase transition~\cite{Endrodi:2011gv,ref:LQCDCplot,ref:lqcd03,ref:lqcd05}. If this prediction holds true, a critical point must exist on the phase diagram, marking the endpoint of the first-order phase boundary.
  Despite ongoing theoretical discussions on the location and even the existence of the QCD critical point, the
  relativistic heavy-ion collisions, by recreating the extreme conditions required to probe the QCD phase structure, provide a experimental platform to search for this critical point~\cite{ref:SPS_result,ref:RHIC_result,Gupta:2011wh,Luo:2017faz,Bzdak:2019pkr,Aggarwal:2010wy,Luo:2015doi,Adamczyk:2014fia,Adamczyk:2013dal,Adamczyk:2017wsl,Adam:2020unf}.
  As the system approaches the QCD critical point, the correlation length of fluctuations grows, leading to enhanced density fluctuations. Conserved quantities such as net-baryon, net-charge, and net-strangeness exhibit fluctuations that are sensitive to the correlation length. The STAR experiment has conducted extensive measurements of high-order cumulants and second-order off-diagonal cumulants of net-proton, net-charge, and net-kaon multiplicity distributions~\cite{RN182,RN183,RN184,RN185}. Notably, a non-monotonic behavior in the fourth-order net-proton cumulant ratio was observed, with a minimum around 19.6 GeV in Au+Au collisions spanning a broad energy range ($\sqrt{s_{\mathrm{NN}}} = 7.7$-200 GeV). This behavior cannot be accounted for by existing theoretical models without invoking the physics associated with the QCD critical point~\cite{Luo:2017faz}.

  In addition to cumulant measurements, baryon density fluctuations arising from critical phenomena during a first-order phase transition are predicted to influence the production of light nuclei~\cite{Sun:2017xrx,Deng:2020zxo,Shuryak:2019ikv,Yu:2018kvh,Shao:2019xpj,RN9}. For example, the light nuclei yield ratio, $N_p N_t / N_d^2$, involving protons $(p)$, deuterons $(d)$, and tritons $(t)$, can be related to the relative neutron density fluctuation, $\langle (\delta n)^2\rangle / \langle n\rangle^2$ (denoted as $\Delta n$ in Ref.~\cite{Sun:2017xrx}). Measurements by the STAR experiment revealed a clear non-monotonic energy dependence of this yield ratio in central Au+Au collisions, with a peak observed at $\sqrt{s_{\mathrm{NN}}} = 20\sim30$ GeV~\cite{Zhang:2019wun,Zhang:2020ewj}.

  Our previous work~\cite{RN2023}, based on the PYTHIA8 Angantyr model, investigated the role of two-body neutron-proton density correlations $(C_{np})$ in light nuclei yield ratios. However, a critical factor was not considered: the three-nucleon correlation $(C_{n^2p})$ involving two neutrons and one proton, which directly impacts triton production and, consequently, the overall yield ratio. Given that tritons form via coalescence processes requiring multiple nucleons, incorporating $C_{n^2p}$ is essential to comprehensively understand light nuclei formation dynamics and extract relative neutron density fluctuations from experimental data.

  In this study, we expand on our earlier findings by incorporating the three-nucleon correlation $C_{n^2p}$ into the analysis of light nuclei yield ratios in Au+Au collisions. Using the PYTHIA8 Angantyr model, we simulated collisions at $\sqrt{s_{\mathrm{NN}}} = 7.7$, 11.5, 14.5, 19.6, 27, 39, 62.4, and 200 GeV.
  We take the notation $R = N_p N_t / N_d^2$ in this study.
  The Angantyr model extends PYTHIA8 by enabling the construction of heavy-ion collisions as a superposition of binary nucleon-nucleon collisions.
  Our findings highlight the critical importance of $C_{n^2p}$ in interpreting light nuclei yield ratios and neutron density fluctuations. By providing an improved methodology for extracting neutron density fluctuations from experimental data, this study offers new insights into the search for the QCD critical point.

  This paper is organized as follows: Section~\ref{subsec:Angantyr} provides an overview of the PYTHIA8 Angantyr model. Section~\ref{sec:2correl} discusses the relationship between neutron density fluctuations and light nuclei yield ratios in heavy-ion collisions. Results and discussions on neutron density fluctuations and neutron-proton correlations are presented in Section~\ref{sec:result}. Finally, we summarize our findings and conclusions in Section~\ref{sec:summary}.

\section{Event generation and definition of light nuclei yield ratio}
\label{sec:Generation_Methodology}

\subsection{PYTHIA8 (Angantyr) model}
\label{subsec:Angantyr}

  PYTHIA~\cite{Sjostrand:2006za} is a extensively utilized event generator designed for simulating particle collisions, particularly proton-proton (pp) and proton-lepton interactions. In pp collisions, Multi-Parton Interaction (MPI) is generated under the assumption that every partonic interaction is almost independent. However, in its default configuration, PYTHIA8 does not natively support heavy-ion collision simulations.

   To overcome this limitation, the PYTHIA8 Angantyr model~\cite{Bierlich:2018xfw} was introduced. This extension enables the extrapolation of pp dynamics into heavy-ion collisions, facilitating the study of proton-nucleus (pA) and nucleus-nucleus (AA) interactions. The Angantyr model achieves this by combining multiple nucleon-nucleon collisions to simulate a single heavy-ion collision. It incorporates theoretical frameworks to describe both hard and soft interactions, along with key features such as initial- and final-state parton showers, particle fragmentation, MPI, color reconnection (CR) mechanisms. Notably, the Angantyr model does not include mechanisms to account for the formation of the quark-gluon plasma (QGP), which is widely believed to occur in AA collisions.

   In the current version of the PYTHIA8 Angantyr model~\cite{Sjostrand:2008za}, heavy-ion collisions are modeled using the Glauber approach to determine the number of participating nucleons based on the geometric overlap of the colliding nuclei. The model introduces algorithms to distinguish between different types of nucleon-nucleon interactions, such as elastic, diffractive, and absorptive processes. It has been shown to describe final-state observables effectively, including particle multiplicity distributions and transverse momentum spectra in AA collisions~\cite{Abdel:2022,Samsul:2022}.

   In this work, we utilized version 8.308 of PYTHIA8 for the simulation of Au+Au collisions at various energies. Approximately one million events were generated for each collision energy, employing multiple PYTHIA tunes with distinct configurations for MPI and CR.
   To remove the influence of system volume and density fluctuations inherent in heavy-ion collisions, we introduced two dimensionless statistical quantities: $\langle (\delta p)\rangle /\langle p\rangle$ and $\langle (\delta n)\rangle /\langle n\rangle$ (described in detail in Sec.~\ref{sec:2correl}). Nucleons were analyzed across various rapidity ranges and centrality classes. The centrality intervals were defined based on the
    transverse momentum sum of charged particles ($\sum E_T$) at the final state of the collisions
   within the pseudorapidity range of $[-0.5, 0.5]$.

  \subsection{Definition of light nuclei yield ratio with three-nucleon correlations}
  \label{sec:2correl}

  The production of light nuclei, such as deuterons and tritons, has been extensively studied to understand the underlying nucleon coalescence mechanism, which plays a pivotal role in high-energy nuclear collisions~\cite{RN9,RN172,RN2023}. This mechanism assumes that nucleons, protons and neutrons within close spatial and momentum proximity can coalesce to form light nuclei.
  The theoretical yield of light nuclei can be derived by applying the coalescence model, neglecting the effects of binding energy. The expression for the abundance of a cluster containing $A$ nucleons is given by:
  \begin{eqnarray}
  N_c = g_c A^{3/2} \left(\frac{2\pi}{m_0T_{\rm eff}}\right)^{3(A-1)/2} V \nonumber \\
  \times \langle \rho_p \rangle^{A_p} \langle \rho_n \rangle^{A_n} \sum_{i=0}^{A_p} \sum_{j=0}^{A_n} C_{A_p}^i C_{A_n}^j C_{n^j p^i},
  \end{eqnarray}
  where $g_c = \frac{2S+1}{2^A}$ is the coalescence factor for a cluster with $A = A_n + A_p$ nucleons and total spin $S$. The terms $m_0$, $T_{\rm eff}$, and $V$ represent the nucleon mass, effective kinetic freeze-out temperature, and system volume, respectively. $\langle \rho_n \rangle$ and $\langle \rho_p \rangle$ denote the average neutron and proton densities, while $C_{A_p}^i$ and $C_{A_n}^j$ are combinatorial factors representing possible configurations of $A_p$ protons and $A_n$ neutrons. The factor $C_{n^j p^i}$ accounts for the correlation between $i$-protons and $j$-neutrons, defined as:
  \begin{equation}
  C_{n^j p^i} = \frac{\langle \delta \rho_p^i \delta \rho_n^j \rangle}{\langle \rho_p \rangle^i \langle \rho_n \rangle^j},
  \end{equation}
  where $\delta \rho_p^i$ and $\delta \rho_n^j$ represent density fluctuations for protons and neutrons, respectively.

  Among these correlations, the relative neutron density fluctuation $\Delta \rho_n = \sigma_n^2 / \langle \rho_n \rangle^2$ corresponds to $C_{n^2 p^0}$. Similarly, the two-nucleon correlation $C_{np}$ can be expressed as:
  \begin{equation}
  C_{np} = \frac{\langle \delta \rho_p \delta \rho_n \rangle}{\langle \rho_p \rangle \langle \rho_n \rangle} = \frac{\langle \rho_p \rho_n \rangle}{\langle \rho_p \rangle \langle \rho_n \rangle} - 1.
  \label{lnr03}
  \end{equation}
  For higher-order correlations, such as the three-nucleon correlation, $C_{n^2 p}$, the expression becomes:
  \begin{equation}
  C_{n^2 p} = \frac{\langle \delta \rho_p \delta \rho_n^2 \rangle}{\langle \rho_p \rangle \langle \rho_n^2 \rangle} = \frac{\langle \rho_p \rho_n^2 \rangle}{\langle \rho_p \rangle \langle \rho_n \rangle^2} - (1 + 2C_{np}).
  \label{lnr04}
  \end{equation}

  The yields of specific light nuclei can now be explicitly formulated. In Eq.~(\ref{lnr03}) and Eq.~(\ref{lnr04}), the nucleons considered are those initial nucleon.
  For deuterons $( N_d )$ and tritons $( N_t )$, the yields are:
  \begin{eqnarray}
  N_d &=& \frac{3}{2^{1/2}} \left(\frac{2\pi}{m_0T_{\rm eff}}\right)^{3/2} V \langle \rho_p \rangle \langle \rho_n \rangle (1 + C_{np}), \\
  N_t &=& \frac{3^{3/2}}{4} \left(\frac{2\pi}{m_0T_{\rm eff}}\right)^3 V \langle \rho_p \rangle \langle \rho_n \rangle^2 (1 + \Delta \rho_n + 2C_{np} + C_{n^2 p}).
  \end{eqnarray}

  From these equations, the ratio of light nuclei yields, $R$, can be expressed as:
  \begin{equation}
  R = \frac{1}{2\sqrt{3}} \frac{1 + \Delta \rho_n + 2C_{np} + C_{n^2 p}}{(1 + C_{np})^2}.
  \label{lnr}
  \end{equation}

  The term $C_{n^2 p}$, representing three-nucleon correlations, introduces a significant contribution to the light nuclei yield ratio. By assuming $C_{n^2 p} = 0$, the yield ratio simplifies to:
   \begin{equation}
  R = \frac{1}{2\sqrt{3}} \frac{1 + \Delta \rho_n + 2C_{np}}{(1 + C_{np})^2}.
  \label{lnr1}
  \end{equation}
  In the absence of both two- and three-nucleon correlations ($C_{np} = C_{n^2 p} = 0$, the $C_{np}$ and $C_{n^2p}$ are the parameters of nucleon correlation), the expression further simplifies to:
  \begin{equation}
  R = \frac{1 + \Delta \rho_n}{2\sqrt{3}}.
  \label{lnr2}
  \end{equation}

  In this simplified scenario, the yield ratio becomes directly proportional to the relative neutron density fluctuation $\Delta \rho_n$, which forms the experimental basis for extracting the $\Delta \rho_n$ from the yield ratios of light nuclei.

  \section{Results and Discussions}
\label{sec:result}

\begin{figure}[tb]
    \includegraphics[width=0.77\textwidth]{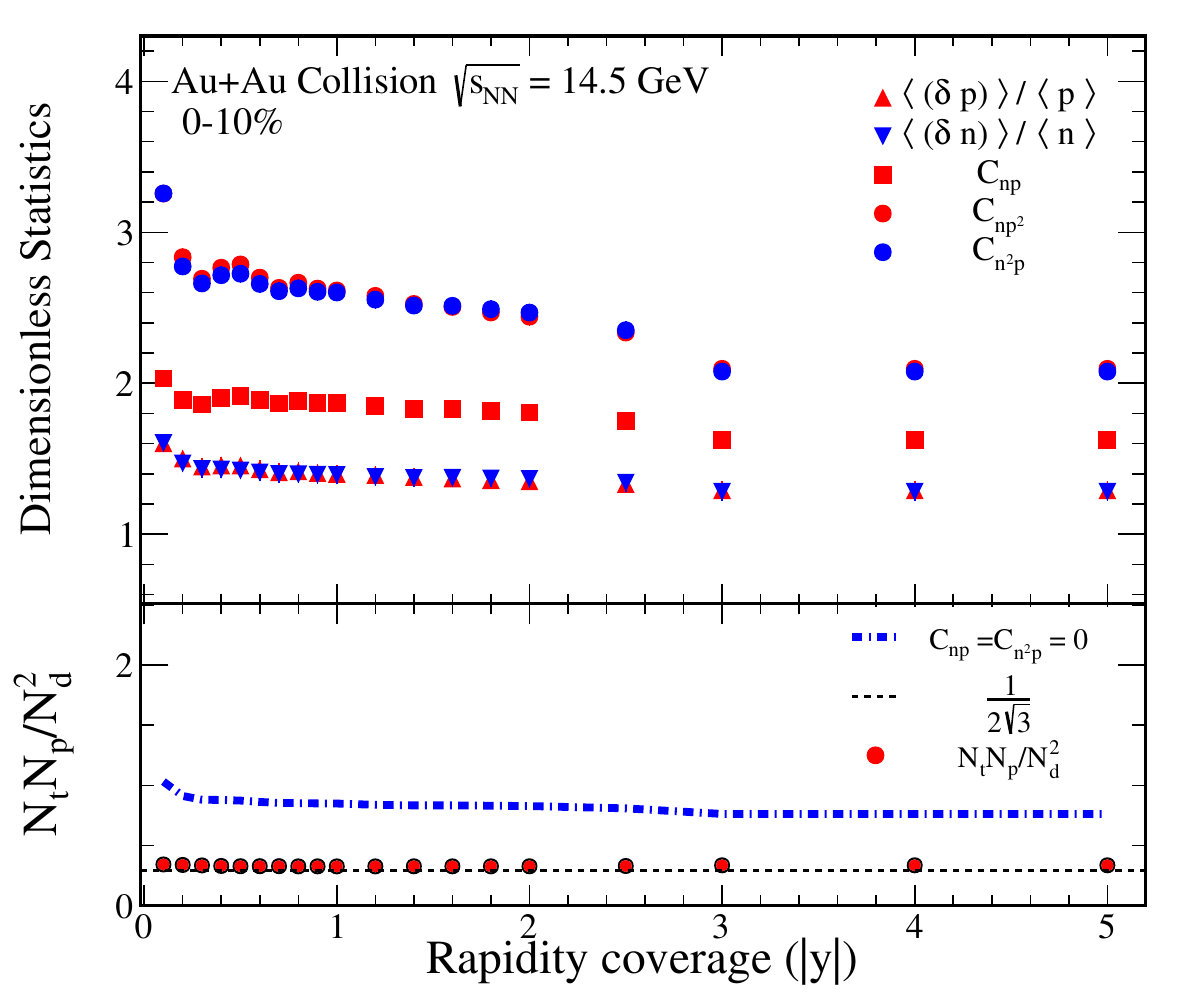}
    \caption{\label{fig1}
    The top panel of the figure present various dimensionless statistical quantities, including $\langle (\delta p) \rangle /\langle p \rangle$, $\langle (\delta n) \rangle /\langle n \rangle$, $C_{np}$, $C_{np^2}$, and $C_{n^2 p}$, derived from 0-10\% Au+Au collisions at a center-of-mass energy of $\snn=$ 14.5 GeV. The bottom panel illustrates the yield ratio for light nuclei, $N_tN_p/N_d^2$, which is calculated using the results from the top panels. Solid circles represent values obtained through Eq.~(\ref{lnr}), while dash-dot lines indicate calculations based on Eq.~(\ref{lnr2}).
    }
\end{figure}

  In addition to the physical parameters employed in the model calculations, as detailed in the last paragraph of Sec.~\ref{subsec:Angantyr}, the analysis also incorporates the option to activate or deactivate the multiple-parton interactions (MPI) based color reconnection (CR) mechanism. This can be achieved by enabling or disabling the parameters \texttt{ColourReconnection:reconnect} and \texttt{PartonLevel:MPI} within the PYTHIA8 Angantyr model framework.

\begin{figure}[tb]
    \includegraphics[width=0.77\textwidth]{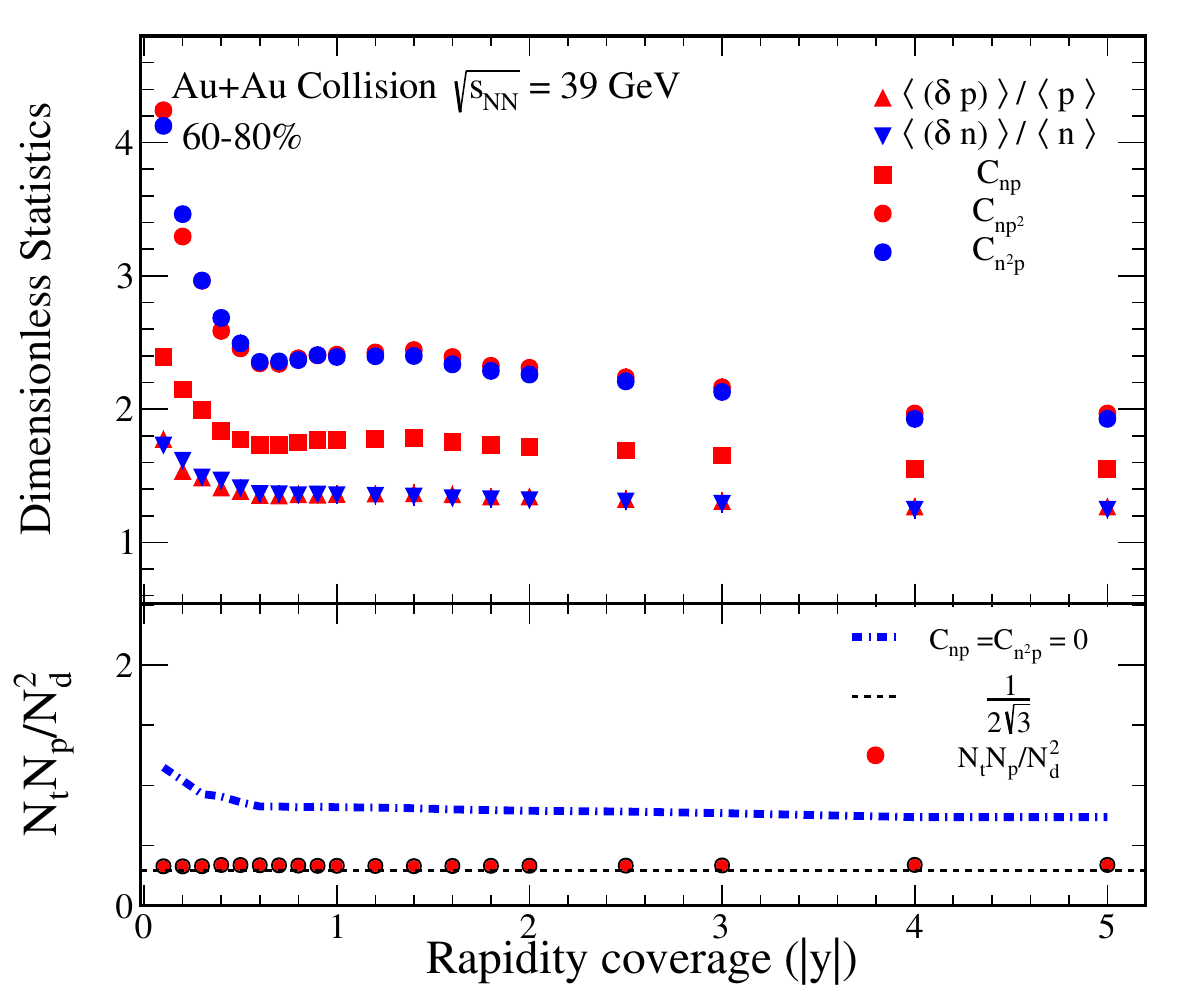}
    \caption{\label{fig2}
  Figure presents various dimensionless statistics and the ratio of light nuclei yields for 60-80\% Au+Au collisions at a center-of-mass energy of $\snn = 39$ GeV. In the top panels, we display the dimensionless statistics $\langle (\delta p)\rangle /\langle p\rangle$, $\langle (\delta n)\rangle /\langle n\rangle$, $C_{np}$, $C_{np^2}$, and $C_{n^2 p}$.
  The bottom panels of the figure illustrate the ratio of yields for light nuclei, given by $N_tN_p/N_d^2$. This ratio is calculated using the data presented in the top panels. The solid circles represent the light nuclei yield ratio, calculated according to Eq.~(\ref{lnr}), while the dash-dot line corresponds to the calculation based on Eq.~(\ref{lnr2}).}
\end{figure}

\begin{figure}[tb]
    \includegraphics[width=0.49\textwidth]{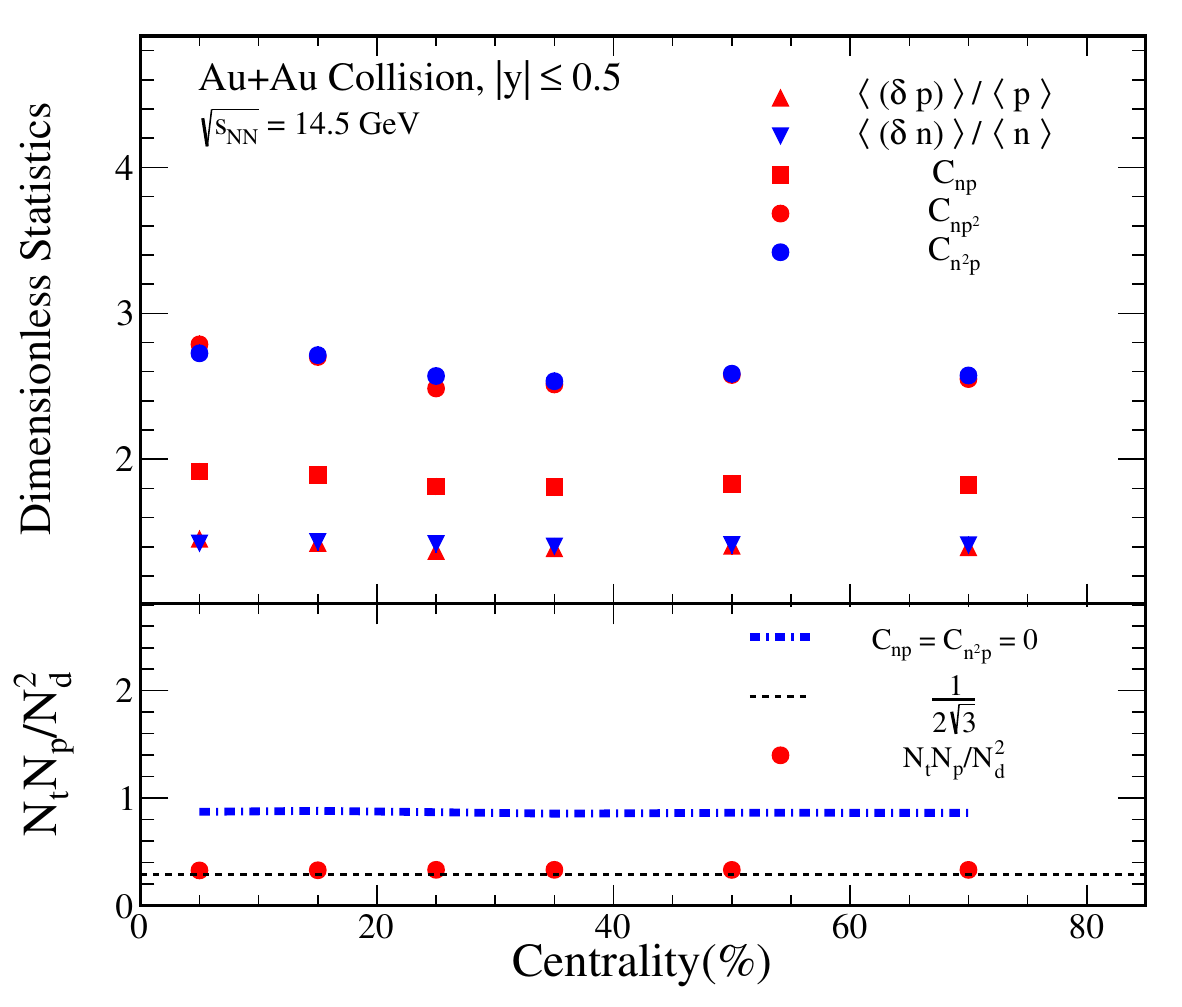}
    \includegraphics[width=0.49\textwidth]{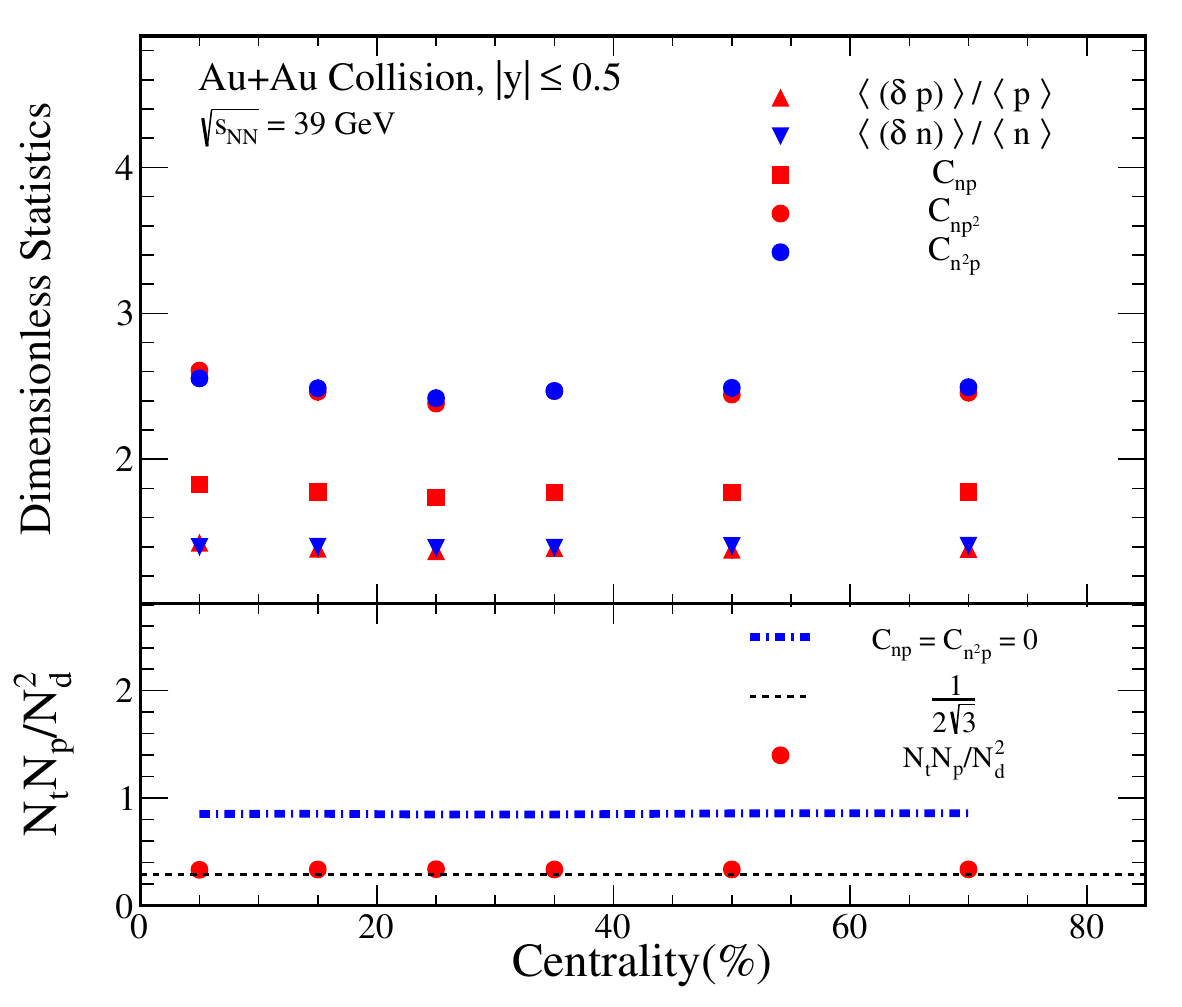}
    \caption{\label{fig3} Figures present the centrality dependent $\langle (\delta p)\rangle /\langle p\rangle$, $\langle (\delta n)\rangle /\langle n\rangle$, $C_{np}$, $C_{np^2}$ and $C_{n^2 p}$ for Au+Au collisions at center-of-mass energies of $\snn=$ 14.5 and 39 GeV, with rapidities confined to the range $|y|\leq 0.5$. Additionally, the ratio of yields for light nuclei, given by $N_tN_p/N_d^2$, is shown in the same figure. This ratio is represented by solid circles, calculated according to Eq.~(\ref{lnr}), while the dash-dot line corresponds to the result obtained from Eq.~(\ref{lnr2}).
    }
\end{figure}

\begin{figure}[tb]
    \includegraphics[width=0.79\textwidth]{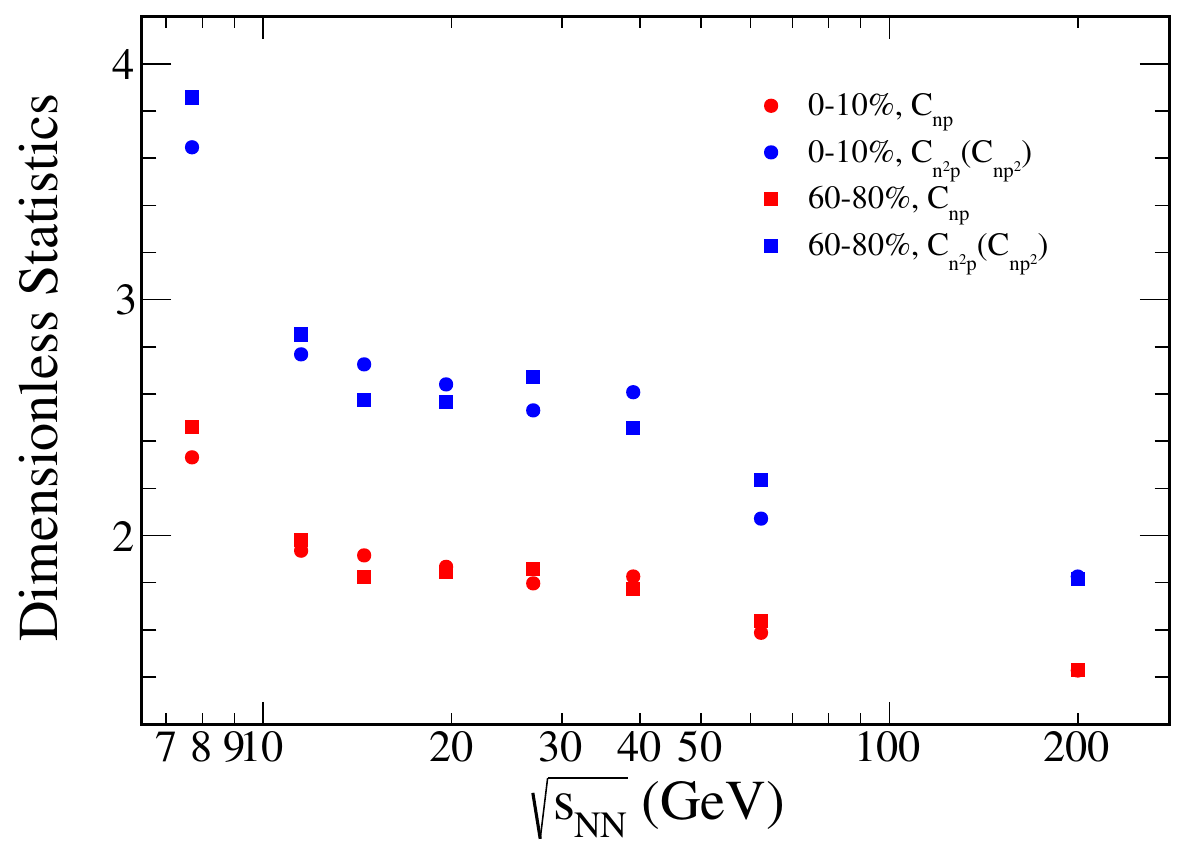}
    \caption{\label{fig4}
     Figures present the collision energy dependence of $C_{np}$, $C_{np^2}$ and $C_{n^2 p}$ for Au+Au collisions, with rapidities confined to the range $|y|\leq 0.5$.  The results for 0-10\% central collisions are represented by solid circles, while those for 60-80\% peripheral collisions are depicted as squares.
    }
\end{figure}

  The results presented in Figures~\ref{fig1}, \ref{fig2}, \ref{fig3}, \ref{fig4}, and \ref{fig6} are obtained by applying the PYTHIA8 Angantyr model, with the inclusion of the MPI-based CR mechanism. These results provide significant insights into the interplay between nucleon density fluctuations and light nuclei yield ratios in high-energy nuclear collisions.

  Focusing on the top panel of Figure~\ref{fig1}, the rapidity dependence of the dimensionless statistics $\langle (\delta p) \rangle /\langle p \rangle$ and $\langle (\delta n) \rangle /\langle n \rangle$ is shown for 0-10\% Au+Au collisions at $\snn=$ 14.5 GeV. As the rapidity coverage increases, both $\langle (\delta p) \rangle /\langle p \rangle$ and $\langle (\delta n) \rangle /\langle n \rangle$ demonstrate a clear decreasing trend, suggesting that relative nucleon density fluctuations reduce with wider rapidity coverage. Furthermore, these fluctuations appear to converge toward a constant value as the rapidity coverage is extended. Notably, the $\langle (\delta p) \rangle /\langle p \rangle$ and $\langle (\delta n) \rangle /\langle n \rangle$ are consistent in Figure~\ref{fig1}.
  The correlations $C_{np}$, $C_{np^2}$, and $C_{n^2 p}$ exhibit a similar trend to the relative nucleon density fluctuations with respect to rapidity coverage for 0-10\% Au+Au collisions at $\snn=$ 14.5 GeV.

  To provide a baseline for comparison, a reference line at $1/2\sqrt{3}$ is plotted, corresponding to the scenario where both density fluctuations and correlations vanish. The light nuclei yield ratio, $N_tN_p/N_d^2$, calculated using Eq.~(\ref{lnr}), is depicted as solid circles in the bottom panel of Figure~\ref{fig1}. This ratio consistently exceeds the reference line of $1/2\sqrt{3}$ for 0-10\% central Au+Au collisions at $\snn = 14.5$ GeV. The results further reveal that when the correlations $C_{np}$ and $C_{n^2 p}$ are neglected (as represented by the dash-dot lines derived from Eq.~(\ref{lnr2})), the calculated yield ratio becomes significantly higher than the reference line. This observation highlights the critical role of $C_{np}$ and $C_{n^2 p}$ in determining the relative neutron density fluctuations through the light nuclei yield ratio.
  Similar trends are observed at other collision energies for central collisions, underscoring the generality of these results.

  Figure~\ref{fig2} presents the rapidity dependence of various dimensionless statistics and the light nuclei yield ratio for 60-80\% Au+Au collisions at a center-of-mass energy of $\snn = 39$ GeV. The top panels display the ratios $\langle (\delta p)\rangle /\langle p\rangle$ and $\langle (\delta n)\rangle /\langle n\rangle$, along with the correlations $C_{np}$, $C_{np^2}$, and $C_{n^2 p}$. In these top panels, it is evident that both $\langle (\delta p)\rangle /\langle p\rangle$ and $\langle (\delta n)\rangle /\langle n\rangle$ decrease as the rapidity coverage increases. Furthermore, the correlations $C_{np}$, $C_{np^2}$, and $C_{n^2 p}$, shown in the same panels, are also dependent on the rapidity coverage. These correlations follow a similar trend to the relative fluctuations in nucleon density.

  In the bottom panels of Figure~\ref{fig2}, we show the ratio of yields for light nuclei, specifically $N_tN_p/N_d^2$. This ratio is calculated using the data from the top panels and is represented by solid circles in accordance with Eq.~(\ref{lnr}), while the dash-dot line corresponds to the calculation using Eq.~(\ref{lnr2}). Our results indicate that the exclusion of the $C_{np}$ and $C_{n^2 p}$ parameters leads to a significant increase in the light nuclei yield ratio.
  This indicates that neglecting the $C_{np}$ and $C_{n^2 p}$ parameters reduces the relative fluctuations in neutron density, thereby resulting in a higher light nuclei yield ratio.

  Figure~\ref{fig3} presents the centrality dependence of the dimensionless statistics $\langle (\delta p)\rangle /\langle p\rangle$ and $\langle (\delta n)\rangle /\langle n\rangle$ for Au+Au collisions at $\snn = 14.5$ and 39 GeV, with rapidity coverage of $|y|\leq 0.5$. As shown in the top panels, both quantities remain flat across central to peripheral collisions.

  In the bottom panels of Figure~\ref{fig3}, the ratios of yields for light nuclei, as defined by Eq.~(\ref{lnr}) and Eq.~(\ref{lnr2}), are also observed to be flat across central, mid-central, and peripheral collisions. However, it is important to note that the relative nucleon density fluctuation cannot be directly extracted from the light nuclei yield ratios alone. The effects of neutron-proton correlations, particularly $C_{np}$ and $C_{n^2 p}$, must be taken into account for a comprehensive understanding of the fluctuation behavior at different centralities.

   Figure~\ref{fig4} illustrates the collision energy dependence of the two- and three-nucleon correlations ($C_{np}$, $C_{np^2}$ and $C_{n^2 p}$) for Au+Au collisions with rapidities confined to $|y|\leq 0.5$, spanning from 0-10\% central to 60-80\% peripheral collisions. As shown in the figure, the
   light nuclei yield ratio exhibits a decrease with rising collision energy.

\begin{figure}[tb]
    \includegraphics[width=0.79\textwidth]{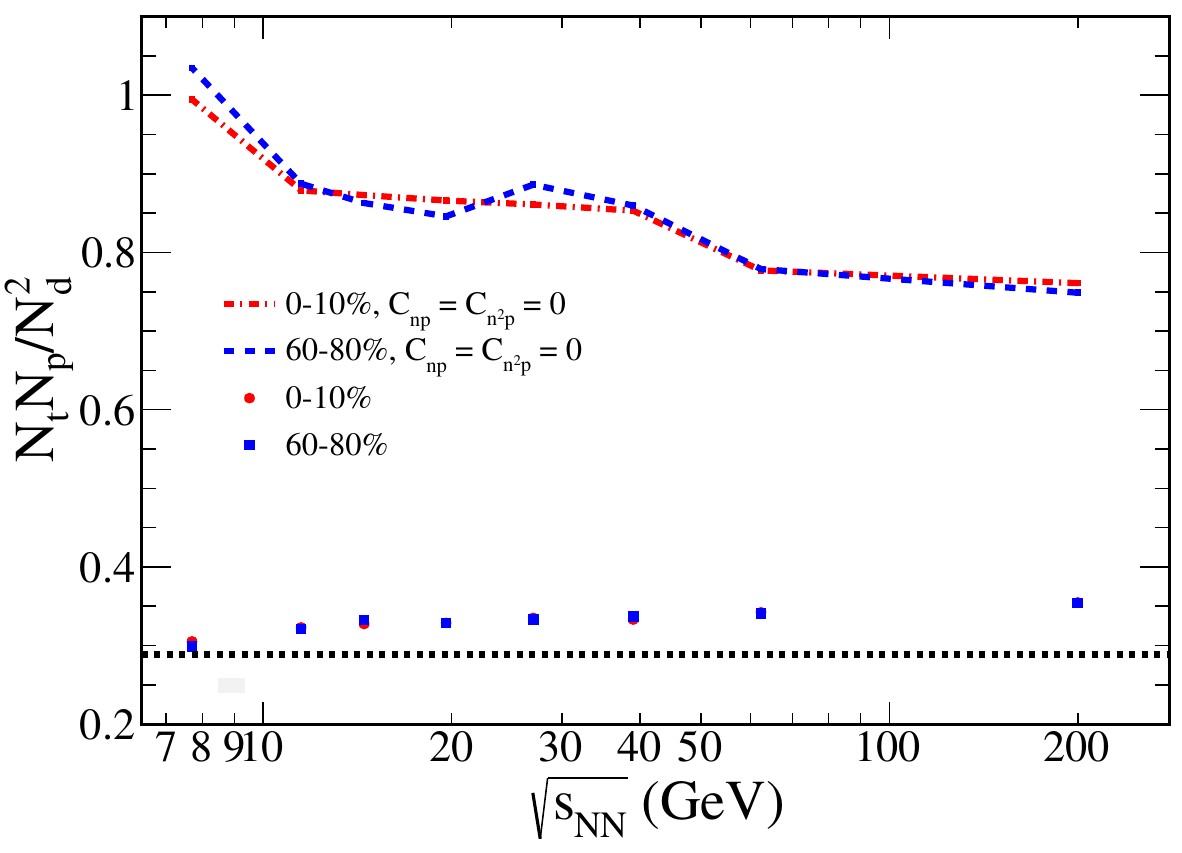}
    \caption{\label{fig5}
    The PYTHIA8 Angantyr model was employed to investigate the collision energy dependence of the light nuclei yield ratio, $N_tN_p/N_d^2$, in Au+Au collisions with rapidities restricted to $|y|\leq 0.5$. The results for 0-10\% central collisions are represented by solid circles, while those for 60-80\% peripheral collisions are depicted as squares. Additionally, the dash-dot lines show the corresponding results when the correlations $C_{np}$ and $C_{n^2 p}$ are excluded.
    }
\end{figure}

   Figure~\ref{fig5} illustrates the collision energy dependence of the light nuclei yield ratio, $N_tN_p/N_d^2$, for Au+Au collisions with rapidities confined to $|y|\leq 0.5$, spanning from 0-10\% central to 60-80\% peripheral collisions. As shown in the figure, the light nuclei yield ratio exhibits a slight increase with rising collision energy.

   It is noteworthy that the yield ratio for peripheral collisions is comparable to that of central collisions, with both ratios exceeding the value of $1/2\sqrt{3}$. Additionally, the results for the case where the correlations $C_{np}$ and $C_{n^2 p}$ are excluded are also presented in the figure, indicated by the dash-dot lines. These results reveal a decrease in the light nuclei yield ratio as the collision energy increases.

\begin{figure}[tb]
    \includegraphics[width=0.49\textwidth]{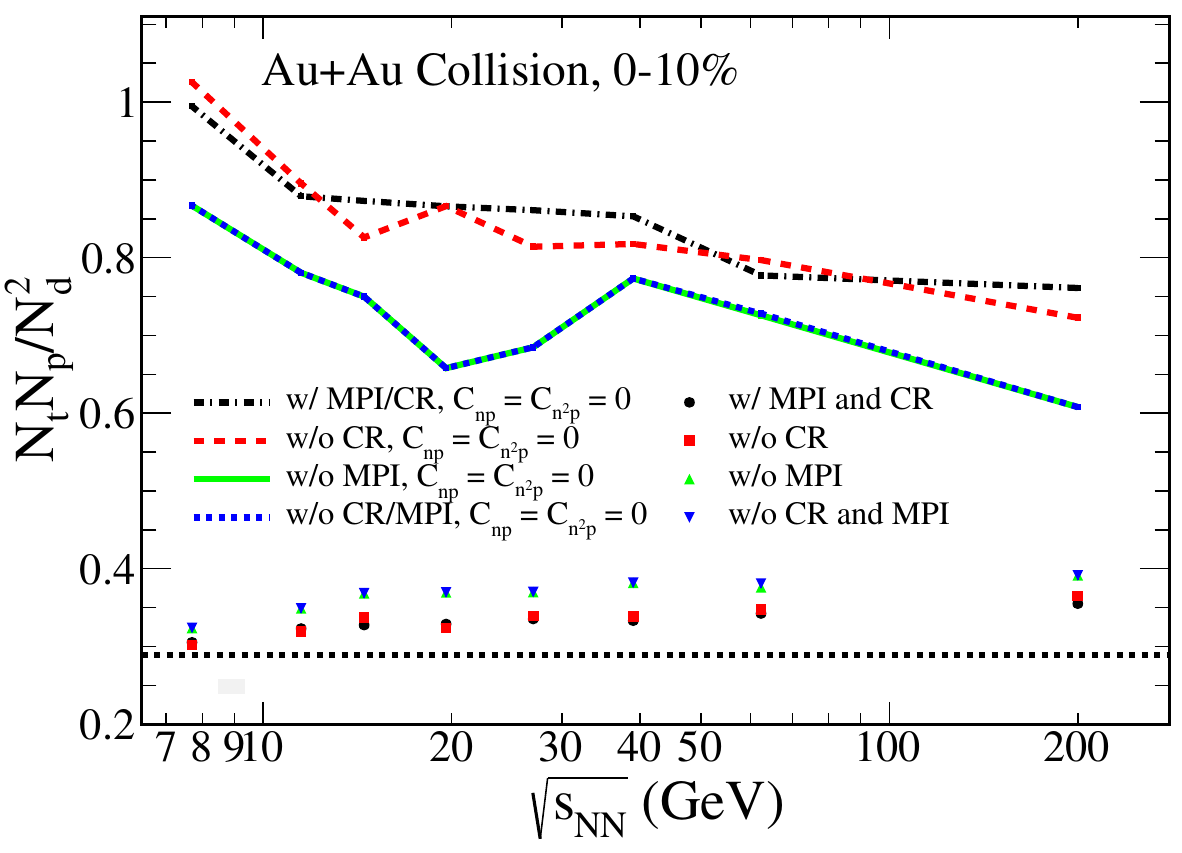}
    \includegraphics[width=0.49\textwidth]{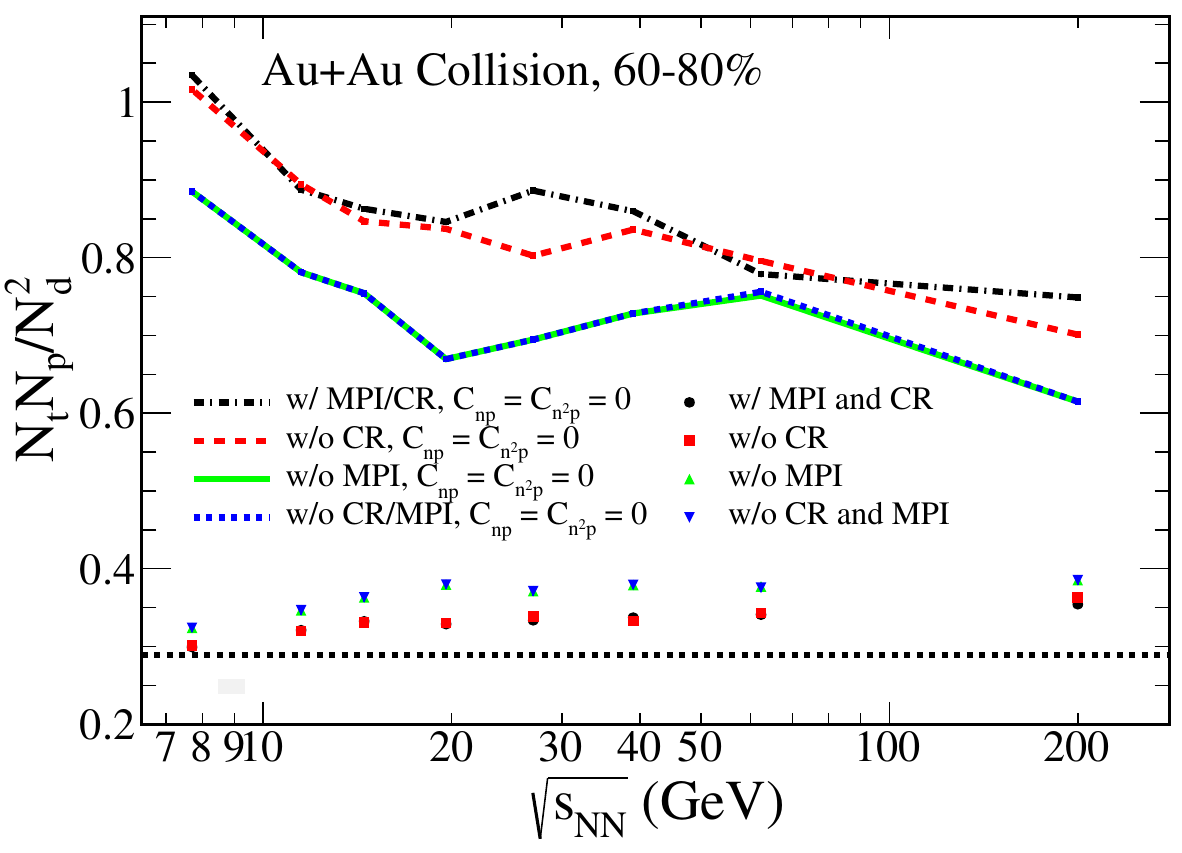}
    \caption{\label{fig6}
    The collision energy dependence of the light nuclei yield ratio $N_tN_p/N_d^2$ in Au+Au collisions with $|y|\leq 0.5$ was investigated using the PYTHIA8 Angantyr model across different model tunes. The results, displayed in the upper and lower panels of Figure~\ref{fig5}, correspond to 0-10\% central and 60-80\% peripheral collisions, respectively.
    }
\end{figure}

  To investigate the effect of different PYTHIA8 Angantyr model tunes, we explore several configurations: MPI with CR, No CR, No MPI, and both MPI and CR turned off. The collision energy dependence of the light nuclei yield ratio $N_tN_p/N_d^2$ in Au+Au collisions with rapidity coverage $|y|\leq 0.5$ is displayed in Figure~\ref{fig6}, which shows results for both 0-10\% central and 60-80\% peripheral collisions.

  From the Figure~\ref{fig6}, it is apparent that the light nuclei yield ratio increases slightly as the collision energy increases across the different configurations of the PYTHIA8 Angantyr model. This trend is observed for both central and peripheral collisions. Furthermore, the results for the cases in which $C_{np}$ and $C_{n^2 p}$ are excluded, represented by dash-dot lines, reveal a decrease in the light nuclei yield ratio as the collision energy increases. In the context of Au+Au collisions, the absence of CR does not produce a significant effect when MPI is turned off, regardless of the PYTHIA8 Angantyr model tune.

  When $C_{n^2p}$ and $C_{np}$ are taken into account, a reduction in the yield ratio is observed for both central and peripheral collisions in Figure~\ref{fig5} and Figure~\ref{fig6}. This suggests that neglecting these correlation terms may lead to an overestimation of neutron density fluctuations . In the lower energy regime, the influence of $C_{n^2p}$ and $C_{np}$ on the yield ratio is significant. However, the underlying mechanisms remain unclear and represent an important direction for future research.

 \begin{figure}[tb]
    \includegraphics[width=0.79\textwidth]{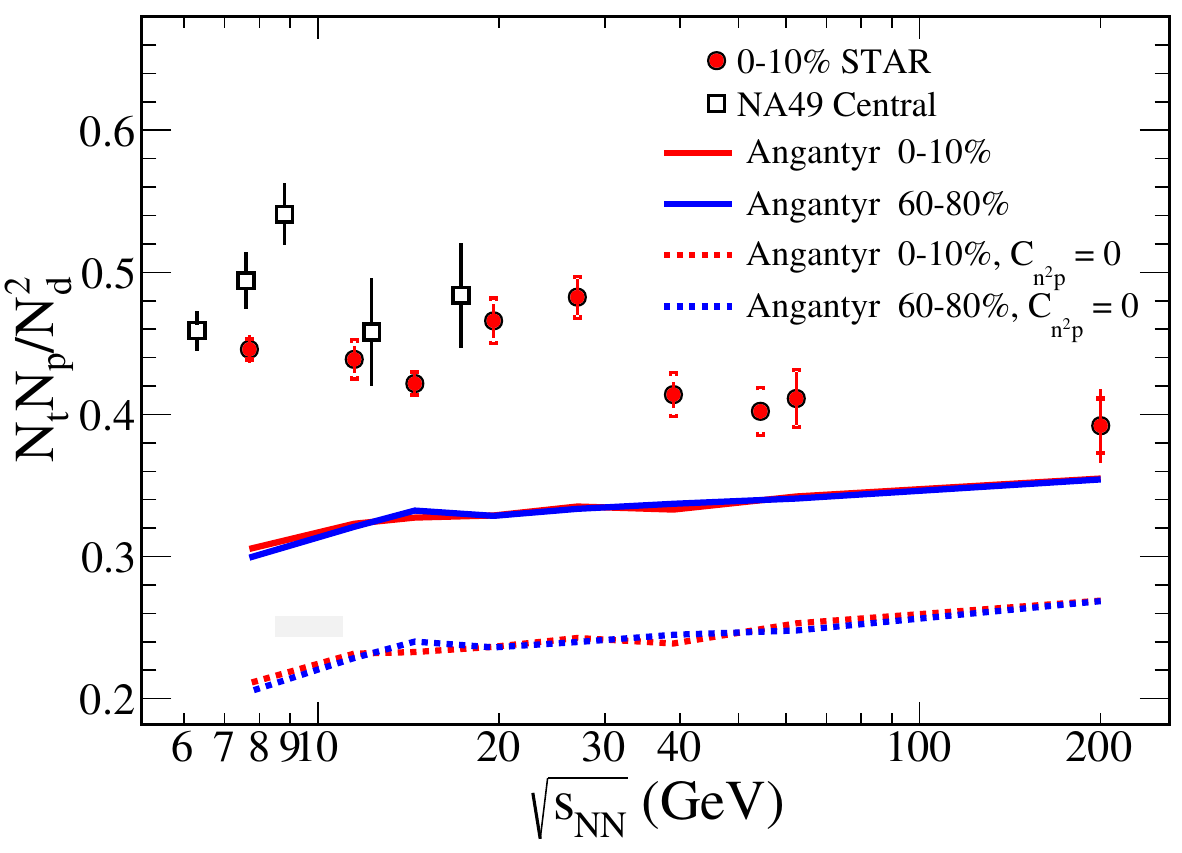}
    \caption{\label{fig7}
     The yield ratio of $N_tN_p/N_d^2$, considering both collision energy and centrality, was analyzed using the PYTHIA8 Angantyr model within the rapidity range $|y|\leq 0.5$. The results from the PYTHIA8 Angantyr model are represented as solid lines under the MPI and RC mode, while dash lines indicate results from the same model with ignore three-nucleon correlation ($C_{n^2p} = 0$), as detailed in~\cite{RN2023}.
     Experimental results from the STAR detector for 0-10\% central Au+Au collisions~\cite{Zhang:2020ewj,2209.08058} are shown as solid circles, while data from the NA49 experiment in central Pb+Pb collisions~\cite{RN9,RN19} are presented as open squares.
    }
\end{figure}

   Figure~\ref{fig7} presents the experimental results of light nuclei yield ratios from the STAR detector in 0-10\% central Au+Au collisions~\cite{Zhang:2020ewj,2209.08058} and NA49 in central Pb+Pb collisions~\cite{RN9,RN19}. These results are compared with theoretical predictions from the PYTHIA8 Angantyr model. A striking feature observed in the experimental data is the non-monotonic energy dependence of the yield ratios. Specifically, the yield ratio of light nuclei exhibits a pronounced peak in the energy range of $\snn = 20 \sim 30$ GeV, which indicates the most significant relative neutron density fluctuations occurring within this range.

   In contrast, the PYTHIA8 Angantyr model underestimates the experimental results across the entire energy range. This discrepancy arises because the PYTHIA8 Angantyr model lacks mechanisms involving critical phenomena and the Quark-Gluon Plasma (QGP) medium, making it unable to reproduce the observed non-monotonic energy dependence.

   Furthermore, Figure~\ref{fig7} highlights the role of three-nucleon correlations, denoted as $C_{n^2p}$, which lead to an overall enhancement of the light nuclei yield ratios, irrespective of whether the collisions are central or peripheral.

   The light nuclei ratios in our results show a weak dependence on collision energy, with a slight increase observed at higher energies.Our results underestimate the experimental results, which include all the real physics, such as the QCD critical point effect around 20 GeV and a possible spinodal effect around 8 GeV~\cite{RN9}. Both of these effects might cause a larger light nuclei ratio at lower collision energies. However, our model currently cannot address these effects. So the discrepancy between our results and the experimental results is large at low energies but small at high energies.

\section{Summary}
\label{sec:summary}

   In summary, this study employs the PYTHIA8 Angantyr model to investigate the dependence of the relative neutron density fluctuation, neutron-proton correlations ($C_{np}$ and $C_{n^2p}$), and the corresponding light nuclei yield ratio ($N_tN_p/N_d^2$) on rapidity, centrality, and collision energy.

   A critical observation from the study is the effect of multi-parton interactions and color reconnection (CR) on light nuclei yield ratios. The analysis demonstrates that color reconnection has no impact on the yield ratios if MPI is turned off. Regardless of whether the collisions are central or peripheral, the three-nucleon correlation, $C_{n^2p}$, leads to a consistent enhancement in the light nuclei yield ratios.

   Comparing the PYTHIA8 Angantyr model predictions with experimental data, the experimental results exhibit a prominent peak in the light nuclei yield ratio at $\snn = 20 \sim 30$ GeV. This peak is indicative of significant fluctuations in the relative neutron density, which are linked to critical phenomena in the collision dynamics. The observed non-monotonic energy dependence in the experimental results is, however, underestimated by the PYTHIA8 Angantyr model. This discrepancy arises due to the model's inability to incorporate critical physics and Quark-Gluon Plasma (QGP) medium effects. Despite its limitations, the PYTHIA8 Angantyr model serves as a useful baseline for scenarios where critical phenomena and QGP medium mechanisms are absent.
   Our study provides a baseline for understanding nucleon coalescence effects in the absence of a QGP phase. As such, it should serve as a useful reference for models that incorporate critical dynamics.

  The authors appreciate the referee for his/her careful reading of the paper and valuable comments. This work is supported in part by the Key Laboratory of Quark and Lepton Physics (MOE) in Central China Normal University (No. QLPL2024P01), the China Scholarship Council (No. 202408420279), the NSFC Key Grant 12061141008, and the Scientific Research Foundation of Hubei University of Education
  for Talent Introduction (No. ESRC20230002).


\end{document}